\documentclass[12pt]{iopart}

\usepackage{balance}
\usepackage{times,mathptmx}
\usepackage{graphicx} 

\expandafter\let\csname equation*\endcsname\relax
\expandafter\let\csname endequation*\endcsname\relax
\usepackage{amsmath}
\usepackage{bm}
\usepackage{ulem} 

\usepackage{balance}

\usepackage{color}
 
\begin{document}

\title[Investigation of bi-particle states in gate-array-controlled quantum-dot systems ...]{Investigation of bi-particle states in gate-array-controlled quantum-dot systems aided by machine learning techniques}

\author{G.A. Nemnes\textit{$^{*a,b,c}$}, T.L. Mitran\textit{$^{a,c}$}, A.T. Preda\textit{$^{a,b,c}$}, I. Ghiu\textit{$^{a,b}$}, M. Marciu\textit{$^{b}$} and A. Manolescu\textit{$^{d}$} }
\address{$^{a}$~Research Institute of the University of Bucharest (ICUB), Mihail Kogalniceanu Blvd 36-46, 050107 Bucharest, Romania}
\address{$^{b}$~University of Bucharest, Faculty of Physics, 077125 Magurele-Ilfov, Romania.}
\address{$^{c}$~Horia Hulubei National Institute for Physics and Nuclear Engineering, 077126 Magurele-Ilfov, Romania.}
\address{$^{d}$~Department of Engineering, Reykjavik University, Menntavegur 1, IS-102 Reykjavik, Iceland}

\ead{nemnes@solid.fizica.unibuc.ro}


\begin{abstract}
	Quantum computing architectures require an accurate and efficient description in terms of many-electron states. Recent implementations include quantum dot arrays, where the ground state of a multi q-bit system can be altered by voltages applied to the top gates. An extensive investigation concerning the spectra of the many-electron systems under multiple operation conditions set by external voltages typically requires a relatively large number of Hamiltonian diagonalizations, where the Coulomb interaction is considered in an exact manner. Instead of making exhaustive calculations using high throughput computing, we approach this problem by augmenting numerical diagonalizations with machine learning techniques designed to predict the many-electron eigenvalues and eigenfunctions. To this end, we employ and compare the results from linear regression methods such as multivariate least squares (MLS) as well as non-linear techniques based on kernel ridge regression (KRR), Gaussian process regression (GPR) and artificial neural networks (ANNs). The input feature vectors are assembled from readily available information comprised from a binary representation of the potential and the strength of the Coulomb interaction. Furthermore, employing a linear classifier, we establish a rule for detecting a singlet-triplet transition which may arise for certain potential configurations.
\end{abstract}

%

%
%
%
%

\section{Introduction}

The continuous downscale of electronic devices is motivated by higher switching speeds and packing densities, while new quantum computing architectures have been proposed, which rely on the ability to maintain the wavefunction coherence and perform operations onto the quantum many-body states. Quantum dot (QD) arrays implemented as cross-bar schemes \cite{doi:10.1126/sciadv.aar3960,doi:10.1063/1.5141522} have been experimentally investigated as prototypes which can integrate a large number of q-bits. Understanding the operation of multi-electron devices requires a significant computational effort due to the complexity of the many-body wavefunction. In many cases, the few-electron quantum system of interest is required to be investigated for multiple input conditions, as well as for a range of structural and compositional configurations, which results in a large set of systems with overlapping properties. 

Since the early stages of nanotechnology development, the physics of many-electron systems was investigated using various simulation methods. Many-electron states have been analyzed in QDs using the exact diagonalization method (EDM) using a realistic three dimensional confinement potential \cite{PhysRevB.61.4718}. Spin-density-functional theory and Monte Carlo methods were employed \cite{PhysRevB.67.235307}, discussing geometry effects in rectangular QDs. Other approaches involve effective charge-spin models for QDs, using higher order perturbation theory and WKB approximation, based on a lattice description \cite{PhysRevB.54.4936}.  

In recent years, several machine learning (ML) techniques have been developed as investigation tools for many-body systems, while most of the works were focused on spin models. These include exact representations of many-body wavefunctions \cite{carleo2017} and interactions \cite{rrapaj2021} using restricted Boltzmann machines (RBM), which were used for the detection of Bell non-locality \cite{deng2018} and for describing dynamical properties in Heisenberg spin models \cite{hendry2019}.
Artificial neural networks (ANNs) have been used for entanglement measurement \cite{gray2018}, for identifying the transition between the thermal phase and many-body localized phase in spin systems \cite{wenjia2018} and for the prediction of ground states in the Bose-Hubbard model \cite{saito2018}. Moreover, the ANNs have been used to construct an exact functional for the Hubbard model \cite{PhysRevB.99.075132}, satisfying the Hohenberg-Kohn theorems. Convolutional neural networks were employed for classifying phases in spin systems \cite{Carrasquilla2017}, providing alternatives to computationally demanding Monte Carlo simulations. 

Here, we consider a two-dimensional finite 2-electron system, with an array of top gates, which controls the electrostatic potential below, in the device active region. The voltage corresponding to each gate can be individually set by one of the two possible values, labeled as low and high voltage, yielding an exponential number of potential configurations. The energy spectra of the many-body Hamiltonian are obtained using EDM \cite{PhysRevB.61.4718,PhysRevB.81.155442,PhysRevB.84.115311}, which are essential for the design of optoelectronic nanodevices. Furthermore, the singlet and triplet states can be tuned by the configuration of top gate voltages and the magnitude of the Coulomb interaction. This implies that for the ground state, the total spin of the 2-electron system can be switched from 0 to 1 by certain top gate voltage configurations. 

In fact, singlet-triplet transitions have been widely studied in quantum dot systems, due to their relevance in the fields of quantum simulation and quantum information \cite{barthelemy}. They occur in isolated quantum dots with two confined electrons in the presence of a perpendicular magnetic field \cite{PhysRevB.45.1951} and also in double quantum dots, where the transition can be electrically controlled \cite{hanson2007}. The same type of transition was studied in double quantum dots realized in a 2D topological insulator \cite{sablikov2018}. In multielectronic quantum dots, the transition can be induced by slightly changing the gate voltage, without any applied magnetic field \cite{martins2017, malinowski2018}. Great interest is also paid to singlet-triplet qubits, which can be controlled by the exchange interaction and are considered a step towards scalable quantum computing \cite{kestner2013}.

Solving exhaustively the set of systems by EDM becomes prohibitive even for a moderate number of gates, due to the exponential number of potential configurations. The computational load is further increased if the Coulomb interaction is tuned by the underlying materials, by means of relative electrical permittivity. Therefore, we investigate in how far machine learning techniques can provide a more efficient and still accurate description of the many-body energy spectra. To this end, we employ multi-target regression methods, starting with linear models like multivariate least squares (MLS), followed by non-linear methods like kernel ridge regression (KRR), Gaussian process regression (GPR) and artificial neural networks (ANNs). Moreover, the transition between the singlet and triplet states, evidenced while tuning the Coulomb interaction for some potential configurations, was also identified by a classification algorithm. The ML methods provide a significant reduction of the computational effort for a large set of many-body systems within a given class, being an efficient approach for exploratory calculations.

\section{Model systems and problem formulation}

\begin{figure}[t]
\centering
\includegraphics[width=6.0cm]{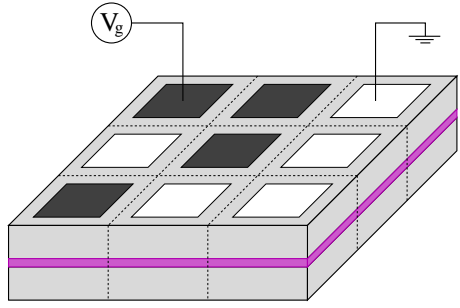}
	\caption{Schematics of the gate array controlling the many-body states in the two-dimensional active region, of size $L \times L$. The gates can be controlled individually, with two different voltages, $V_{\rm g}<0$ and $V_{\rm g}=0$, yielding a total number of $2^{N_{\rm g}}$ potential configurations.}
\label{gatearray}
\end{figure}

We consider a set of $N$-particle quantum systems $(N=2)$ defined on a finite 2-dimensional square shaped region as depicted in Fig.\ \ref{gatearray}. The many-body eigenstates are determined by the voltages applied on the top gates, which form an array with $N_{\rm g} = N_{\rm gx} \times N_{\rm gy}$ elements. We assume each gate voltage can take two values, 0 or $V_{\rm g}$, defining a potential configuration $i_V$, which controls the potential energy $\{V_{xy}(i_V)\}$ in the plane where the electrons are confined. Within this assumption, the main features of the many-electron states are still captured. The model can be further refined, from a more realistic perspective by considering additional screening effects due to the gate electrodes \cite{PhysRevB.61.4718} and local potentials \cite{doi:10.1063/1.359446}. However, this would bring additional complexity, while the main focus is to provide a description of many-electron states, like energy spectra and singlet-triplet transitions using the ML techniques. The Coulomb interaction between electrons is fully accounted for.

The number of systems increases exponentially with the number of top gates, which is $2^{N_{\rm g}}$. Moreover, the relative strength of the Coulomb interaction can be adjusted by the interplay between the dielectric properties of the medium (relative permittivity $\bar{\epsilon}_r$), effective mass $m_{\rm eff}$ and geometrical confinement, of linear size $L$. Other external conditions can be applied, like the in-plane electric fields and/or magnetic fields, to include additional degrees of freedom, leading to a huge number of $N$-particle problems and an exhaustive investigation becomes unfeasible.

\section{Methods}

\subsection{The exact diagonalization method (EDM)}

The Coulomb interaction in the $N$-particle system is considered in an exact manner, which allows a detailed description of the entanglement in the quantum system, as opposed to mean field theories, which is essential for quantum computing applications. The eigensystem is determined using matrix diagonalization in the Fock space \cite{PhysRevB.81.155442,PhysRevB.84.115311}. In the following we detail this procedure, commenting also on the numerical implementation for the given class of many-body systems.

The many-body Hamiltonian in the second quantization is written as:
\begin{equation}
	{\mathcal H} = {\mathcal H_0} + {\mathcal H_{\rm int}} = \sum_a \epsilon_a c_a^\dagger c_a + \frac{1}{2}\sum_{abcd} V_{abcd} c_a^\dagger c_b^\dagger c_d c_c,	
\end{equation}	
where ${\mathcal H_0}$ represents the non-interacting part, which includes the kinetic energy and the external potential $V_{xy}(i_V)$, while ${\mathcal H_{\rm int}}$ accounts for the electron-electron interaction. 

The non-interacting term yields the single-particle solutions, which are given by the single-particle energies $\epsilon_a$ and the spin-dependent functions
\begin{equation}
	|\Phi_a\rangle = \sum_{\sigma_z} \phi_{a,\sigma_z}({\bf r}) |\sigma_z\rangle,
\end{equation}
where $\phi_{a,\sigma_z}({\bf r})$ is the orbital component and $|\sigma_z\rangle$ is an eigenvector of the spin projection along the $z$ axis. These solutions must obey the boundary conditions, which here correspond to vanishing wavefunctions. In order to properly order the spin-states, a small Zeeman term, ${\mathcal H}_{\rm B} = \frac{1}{2} g_{\rm eff} \mu_{\rm B} B_z \sigma_z$, is added to ${\mathcal H_0}$. Here, $g_{\rm eff}$ is the effective gyromagnetic factor of the active material and $\mu_{\rm B}$ is the Bohr magneton.

The Coulomb potential can be adjusted by the relative permittivity $\bar{\epsilon}_r$, which is a material dependent parameter
\begin{equation}
	V_{\rm C}({\bf r-r'}) = \frac{\kappa e^2}{|{\bf r-r'}|}, \;\;\; \kappa = \frac{1}{4\pi\bar{\epsilon}_0\bar{\epsilon}_r}	.
\end{equation}
The maximum Coulomb interaction ($V_{\rm C}^0$) is obtained for $\bar{\epsilon}_r=1$, so that $V_{\rm C} = V_{\rm C}^0 / \bar{\epsilon}_r$.  In general, the impact of the Coulomb interaction on the quantum states depends on its strength relative to the single-particle level spacings, which can be further influenced by $L$ and $m_{\rm eff}$.    

The matrix elements of the Coulomb potential are calculated based on the single-particle functions
\begin{equation}
	V_{abcd} = \langle\Phi_a({\bf r})\Phi_b({\bf r'})|V_{\rm C}({\bf r-r'})|\Phi_c({\bf r})\Phi_d({\bf r'})\rangle ,
\end{equation}
which can be explicitly written as:
\begin{eqnarray}
	V_{abcd} &=& \int d{\bf r} \int d{\bf r'}	\sum_{\sigma_z,\sigma'_z} 
	   \phi^*_{a,\sigma_z}({\bf r}) \phi^*_{b,\sigma_z}({\bf r'}) \times  \nonumber \\ 
	 & & \times \frac{\kappa e^2}{|{\bf r-r'}|}  
           \phi_{c,\sigma_z}({\bf r}) \phi_{d,\sigma_z}({\bf r'}) .
\label{Vabcd}	   
\end{eqnarray}

Having the $V_{abcd}$ matrix elements determined, one may proceed with $N$-particle Hamiltonian diagonalization:
\begin{equation}
{\mathcal H}\Psi_n = E_n \Psi_n, 
\end{equation}
where $E_n$ and $\Psi_n \equiv \Psi_n({\bf r_1, \sigma_{z1}, \ldots, r_N, \sigma_{zN}})$ 
are the eigenvalues and eigenfunctions of the many-body system.

\subsection{Numerical implementation of EDM}

The first step is the numerical diagonalization of the single-particle Hamiltonian, ${\mathcal H_0}$. To this end, we define a 2-dimensional 1-particle basis with spin, which fulfills the boundary conditions:
\begin{equation}
\varphi_{k,\sigma_z}(x,y) = u_i(x) \times u_j(y)\; |\sigma_z\rangle,
\end{equation}
where $u_i(x) = \sqrt{\frac{2}{L}}\sin\left(i\frac{\pi}{L}\left(x+\frac{L}{2}\right)\right)$ and $k=(i,j)$. The number of the basis elements along each of the two spatial directions is $N_{\rm bx}=N_{\rm by}=N_{\rm b}$. Here, the square geometry makes that $x$ and $y$ directions are treated similarly. Therefore, the basis size is $2 \times N_{\rm b}^2$. The matrix elements $\langle\varphi_{k\sigma_z}|{\mathcal H_0}|\varphi_{k'\sigma'_z}\rangle$ are evaluated on a real space grid with $N_x \times N_y$ points. Again, due to the system's symmetry we shall consider $N_x = N_y$.  

The solutions of the 1-particle eigenvalue problem
\begin{equation}
	{\mathcal H_0} \Phi_a(x,y) = \epsilon_a \Phi_a(x,y)
\end{equation}
are $\{\epsilon_a;\Phi_a(x,y)\}$, $a=1,2,\ldots$ in increasing energy order and the eigenfunctions are used to calculate the matrix elements of the Coulomb potential, $V_{abcd}$. The 2-center integrals in Eq.\ (\ref{Vabcd}) are quite computationally demanding even for moderate number of 1-particle states. Therefore, we typically select a small number of single electron states, $N_{\rm SES}$, which define the $N$-particle basis by forming Slater determinants of rank $N$. In the case of Fermions, the number of $N$-particle basis elements is $N_{\rm MES} = C_{N_{\rm SES}}^N$.

The $N$-particle eigenvalue problem is solved using the occupation number representation. The basis elements correspond to the eigenvectors of the non-interacting system and they are represented as binary strings, which indicate the single particle states that form a given Slater determinant. If we take one of these basis elements as a many-electron state denoted by $|{\rm \psi_k}\rangle$, then:
\begin{equation}
	|{\rm \psi_k}\rangle = |n_1^{(k)} n_2^{(k)} \ldots n_s^{(k)} \ldots\rangle, 
\end{equation}
where $n_s^{(k)} = 0$ or $1$ and $s \le N_{\rm SES}$. The action of the creation and annihilation operators is now straightforward to implement \cite{Fetter}:
\begin{eqnarray}
	c_s |\psi_k\rangle &=& (-1)^{S_s(k)} |\ldots n_s^{(k)}-1 \ldots\rangle  \\
	c_s^\dagger |\psi_k\rangle &=& (-1)^{S_s(k)} |\ldots n_s^{(k)}+1 \ldots\rangle,
\end{eqnarray}
if $n_s^{(k)}=1$, otherwise $c_s |\psi_k\rangle = c_s^\dagger |\psi_k\rangle = 0$ for $n_s^{(k)}=0$. Importantly, the sign is set by the phase factor $(-1)^{S_s(k)}$, where $S_s(k) = n_1^{(k)} + n_2^{(k)} + \ldots + n_{s-1}^{(k)}$.

Using the orthogonality of the $N$-particle basis elements, one can calculate the Hamiltonian matrix of the system with Coulomb interactions, which is diagonalized by a specialized LAPACK routine, providing the eigensystem $\{E_n;\Psi_n\}$. This approach, based on second quantization and occupation number representation, has the advantage that the Slater determinants are not directly explicitated, which may become cumbersome for a larger number of electrons in the quantum system (e.g. $N\ge3$). Observables, like charge and spin density in real space, for a certain eigenstate $n$, can be readily obtained:
\begin{equation}
	\bar{\rho}_n({\bf r}) = \sum_k |C_{nk}|^2 \sum_{p=1}^{N} \left[ |\phi_{i_p,\uparrow}|^2 + |\phi_{i_p,\downarrow}|^2 \right]
\label{charge}
\end{equation}	
\begin{equation}
	\bar{\sigma}_{z,n}({\bf r}) = \sum_k |C_{nk}|^2 \sum_{p=1}^{N} \left[ |\phi_{i_p,\uparrow}|^2 - |\phi_{i_p,\downarrow}|^2 \right]
\label{spin}
\end{equation}

The total charge, $Q=Ne$, as well as the total spin of the system, $S_{z,n} = \int d{\rm r} \bar{\sigma}_{z,n}({\bf r})$ are found by integrating Eqs.\ (\ref{charge}) and (\ref{spin}), respectively.

In this way, the EDM is generally formulated for any particle number $N$, being only limited, numerically, by $N_{\rm MES}$.

\subsection{Machine learning techniques}

The problem formulated here has two generic coordinates, defined by the list of potentials $\{V_{xy}(i_V)\}$ and the strength of the Coulomb interaction $v_{\rm C}=V_{\rm C}/V_{\rm C}^0$. The two coordinates are rather different from the perspective of the ML algorithms. The Coulomb interaction, $v_{\rm C}$, can vary continuously, which typically translates into a smooth variation of the eigenvalues $E_n$ and expansion coefficients $C_{nk}$, although for a given state (e.g. ground state), in the case of a singlet-triplet transition, a sharp variation can also be found. The set of $2^{N_{\rm g}}$ potentials are assembled as a second coordinate, using the potential index $i_V = 0, \dots, 2^{\rm N_g} - 1$, which maps the potentials $V_{xy}(i_V)$ as binary strings in an $N_{\rm g}$-dimensional space. Although the elements of this list of potentials are essentially discrete, there is still a good resemblance between different sub-groups.  

The features are constructed as $(N_{\rm g}+1)$-dimensional vectors, containing the potential binary encoding and the Coulomb parameter. This allows a direct connection between the readily available system information and the target quantities. For predicting the whole set of $N_{\rm MES}$ eigenvalues, multi-target regression methods are employed. As a starting point, MLS is considered and implemented by least squares with multiple target output. This method provides an initial assessment, before more complex, non-linear ML approaches are investigated. In this latter class, we employ KRR, GPR and ANNs. 

The MLS, KRR and GPR models are constructed using SciKit Learn \cite{scikitlearn}, while the ANNs are implemented by TensorFlow \cite{TensorFlow} and Keras \cite{Keras} libraries. These methods have intrinsic advantages and limitations, which are outlined in the folowing. The method of linear (or ordinary) least squares consists of fitting the parameters of an overdetermined linear model by minimizing the sum of the squared residuals. MLS can be further constrained by imposing a penalty, known as an L2 norm, for the summed squared magnitudes of the model's coefficients. This is known as ridge regression or Tikhonov regularization. Kernel ridge regression, which combines ridge regression with the kernel trick, consists of learning a nonlinear function by performing linear regression after projecting the data in a high dimensional feature space. By using the kernel trick it is possible to directly compute the inner products, or similarity, between the pairs of data points without explicitly performing the mapping and computing the coordinates in the new, possibly infinite dimensional space. Gaussian process regression is a method of computing distributions over continuous functions that conform to a finite number of observations or measured values. Its name comes from the fact that any finite joint distribution is multivariate Gaussian. On the other hand, artificial neural networks are a type of models inspired by natural neural architectures and operate by using a directed graph structure to compose a large number of simple nonlinear functions. The network as a whole can be tuned to act as a specific function that best fits the data set by adjusting its internal parameters (the weights and biases of the network) using a backpropagated error signal computed by stochastic gradient descent. In contrast to MLS, KRR and GPR, which are all non-parametric models with training and inference computational loads that increase with the dataset, ANNs are parametric models with a fixed number of internal operations, given by the architecture. This parametric aspect of ANNs and the fact that they can be iteratively trained on subsets of the total dataset (using batches) makes them naturally better suited to handle large set sizes.

 The KRR method is implemented using the radial basis function (RBF) kernel and maximum regularization, specified by the parameter $\alpha = 1$. The GPR model uses a similar RBF kernel, with $\sigma_0 =0.5$ and 5 optimization restarts, while the diagonal elements of the kernel matrix are augmented by $\alpha=0.001$. The ANNs are implemented by TensorFlow \cite{TensorFlow} and Keras \cite{Keras} libraries. The ANN architecture is comprised by one hidden layer with 25 neurons, sigmoid activation function, while the learning process was performed with 5000 epochs, with a batch size of 25 examples and a learning rate of 0.001. The loss function is the mean squared error and the Adam optimizer was employed. The train/test accuracies are evaluated using the $R^2$ coefficient of determination.

Furthermore, a classification problem was designed for the identification of the singlet-triplet transition. Here, we used a linear classification approach, which had a comparable performance with other methods in identifying a basic rule from the potential profiles, for which these transitions are likely to occur. 

It is worth mentioning that related implementations of ML techniques, based RBMs can offer a alternatives to finding the ground state of a given many-body Hamiltonian. These provide an internal representation of a many-body state \cite{rrapaj2021,PhysRevB.96.205152}, using a mechanism based on a variational method. A numerical implementation (NetKet) is described by Carleo {\it et al.} \cite{CARLEO2019100311}. However, we adopt here another perspective, which corresponds to learning features out of a relatively wide set of many-electron problems in order to be able to predict the spectrum in new systems from the same class.

More conventional procedures that imply exhaustive calculations of the entire set of systems would lead to proportionally larger times or increased parallel computing resources. Moreover, the computational time associated with the Hamiltonian diagonalization may become significant as it is typically proportional to cube of the basis size, while parallel diagonalization algorithms have limited scalabilities. In contrast, the ML approach takes advantage of the common features that exist in the relatively large set of many-body problems and provide reasonably accurate results, bypassing the diagonalization procedures and Hamiltonian matrix calculations.

\section{Results}

\begin{figure}[t]
\centering
	(a) \hspace*{4cm} (b) \\
\hspace*{0.2cm}\includegraphics[width=4.75cm]{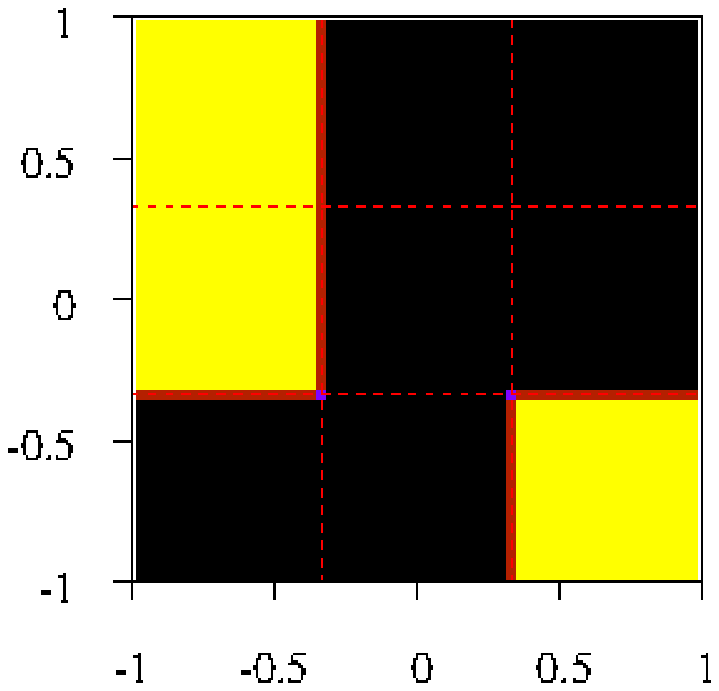} \hspace*{0.6cm}
\includegraphics[width=4.75cm]{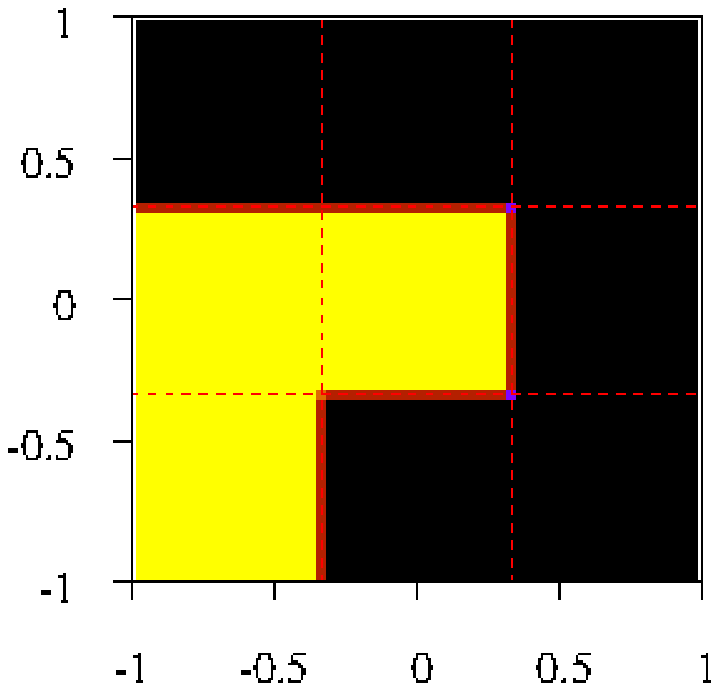} \vspace*{0.5cm}\\
	(c) \hspace*{4cm} (d) \\
\hspace*{0.2cm}\includegraphics[width=4.75cm]{figure_2c} \hspace*{0.5cm}
\includegraphics[width=4.75cm]{figure_2d}
	\caption{Typical potential maps induced by the top-gate array: two potential configurations are represented, ${\mathcal V_1}$ (a) and ${\mathcal V_2}$ (b), corresponding to a system {\it without} and one {\it with} a singlet-triplet transition, respectively. High/low potential energy regions are depicted in yellow/black colors. The unit length is $L/2$. (c,d) The corresponding eigenvalues for different strengths of the Coulomb interaction, $v_{\rm C} = 0, 0.1, 0.5, 1.0$, are represented.}
\label{twosystems}
\end{figure}

\begin{figure}[t]
\centering
\includegraphics[width=3.5cm]{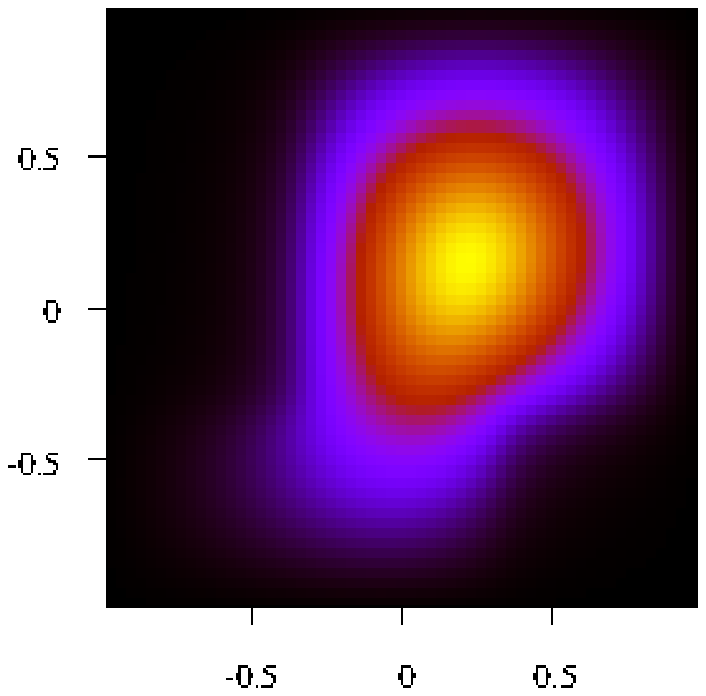} \hspace{0.5cm}
\includegraphics[width=3.5cm]{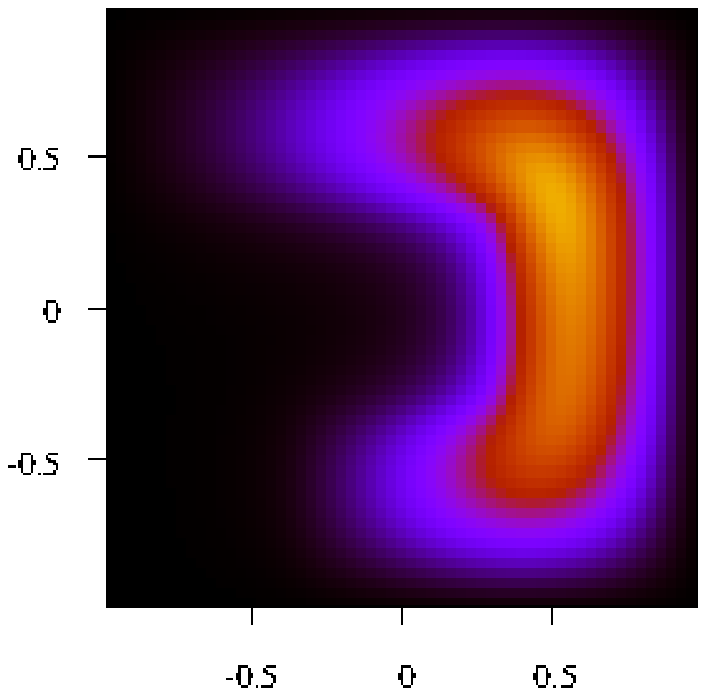} \vspace*{0.2cm}\\
\includegraphics[width=3.5cm]{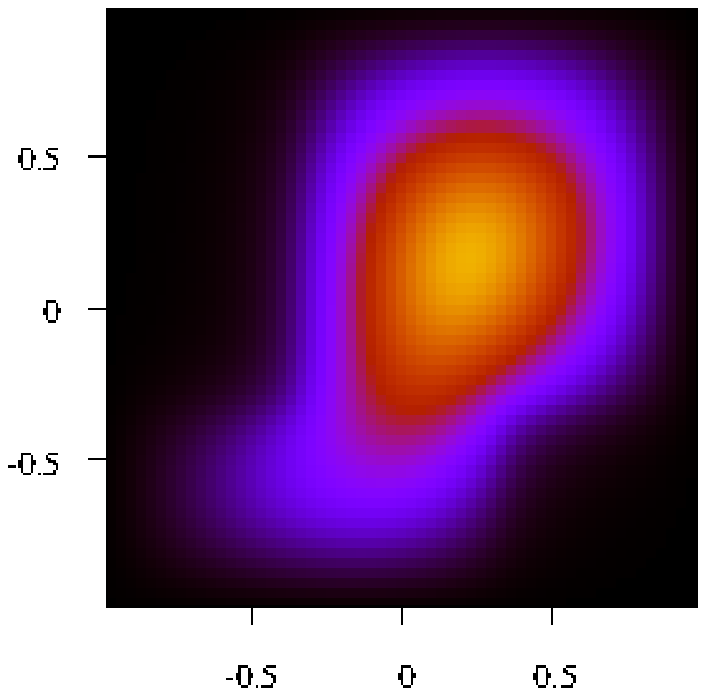} \hspace{0.5cm}
\includegraphics[width=3.5cm]{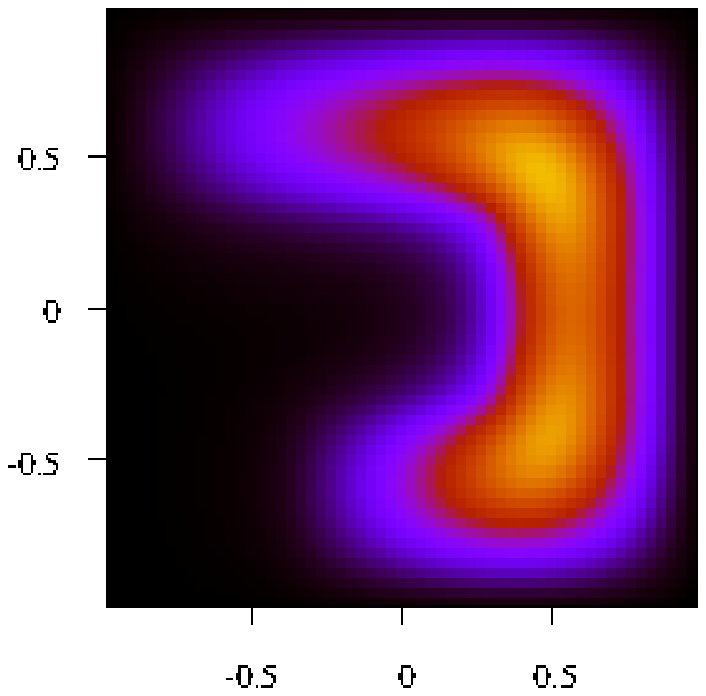} \vspace*{0.2cm}\\
\includegraphics[width=3.5cm]{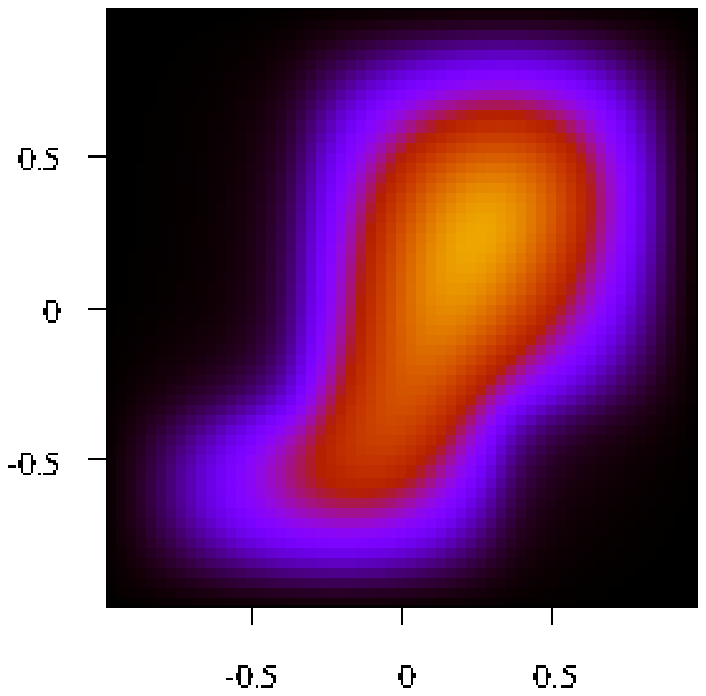} \hspace{0.5cm}
\includegraphics[width=3.5cm]{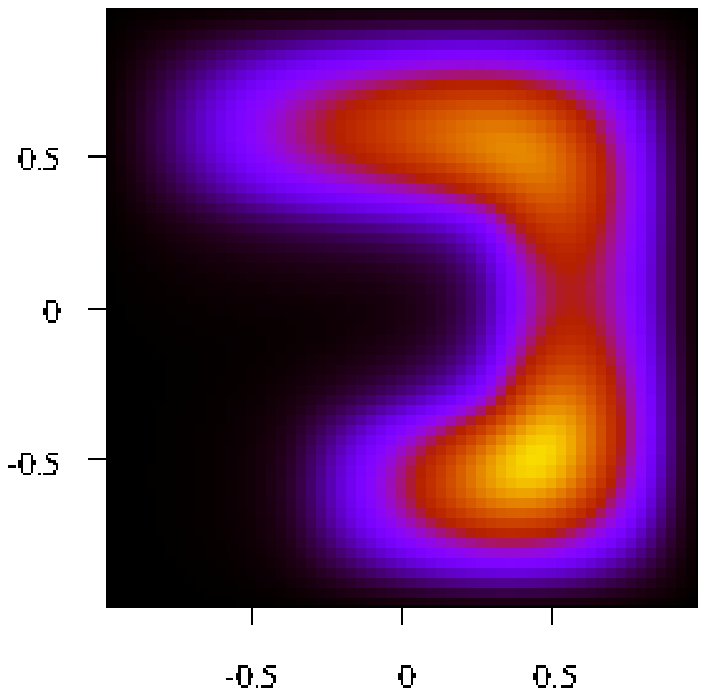} \\
	\caption{Charge densities for the potentials indicated in Fig.\ \ref{twosystems}, ${\mathcal V_1}$ (left column) and ${\mathcal V_2}$ (right column), with varying Coulomb interaction: $v_{\rm C} = 0, 0.5, 1.0$ (from top to bottom). The unit length is $L/2$. The total spin is zero ($S_z=0$) in each instance, except for the case of ${\mathcal V_2}$ with $v_{\rm C} = 1.0$, which corresponds to a singlet-triplet transition.}
\label{chargedensity}
\end{figure}

\begin{figure}[t]
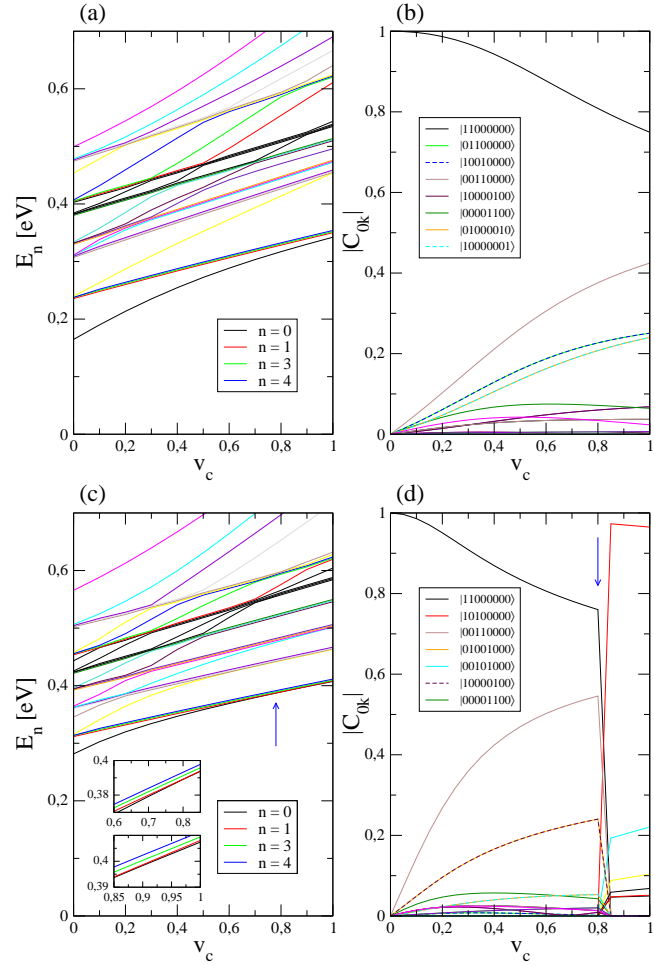

\centering
\includegraphics[width=8.5cm]{figure_4ab} \\
\includegraphics[width=8.5cm]{figure_4cd}
	\caption{Eigenvalues and expansion coefficients $|C_{nk}|$ for potential ${\mathcal V_1}$ (a,b) and potential ${\mathcal V_2}$ (c,d). For the latter case, the vertical arrows mark the transition point for the ground state, corresponding to a change from singlet to triplet state. The individual contributions for several 2-particle basis vectors are indicated. The insets show in more detail the crossing corresponding to the transition point.}
\label{transition}
\end{figure}

In order to evaluate the performance of the ML approaches, we perform exact diagonalizations for all 512 potentials configurations, and for each we consider 21 values for the strength of the Coulomb interaction $v_{\rm C}$, from 0 to 1 in steps of 0.05, totalizing a number of 10752 systems. The single-particle eigenvalue problems are calculated using a basis set of $2 \times N_{\rm bx} \times N_{\rm by}$ functions, with $N_{\rm bx}=N_{\rm by}=30$, on a real space grid with $N_x = N_y = 60$ points. The many electron states are constructed using $N_{\rm SES}=8$ single-electron states with lowest energies, resulting a number of $N_{\rm MES}=28$ bi-particle states. 

In the subsequent calculations we consider the following device parameters: the number of top gates $N_{\rm g}=3^2$, the applied potentials are $V_{\rm g} = 0$ and $0.5$ V, the linear size of the confinement region $L=30$ nm, effective mass $m_{\rm eff}=0.0655$ $m_0$ corresponding to GaAs, where $m_0$ is the mass of the electron in vacuum.

\subsection{Description of the bi-particle states with varying Coulomb interaction}

We first analyze the properties of bi-particle states, by looking at two potential configurations, which are shown to exhibit different behavior when the Coulomb interaction is increased. The selected potentials are depicted in Fig.\ \ref{twosystems}, labeled with ${\mathcal V_1}$ and ${\mathcal V_2}$ and correspond to potential indexes $i_V=70$ and $i_V=19$, respectively. The eigenvalue spectra are indicated for $v_{\rm C} = 0, 0.1, 0.5, 1.0$. For the non-interacting system, if the lowest two single particle states are not degenerated by the orbital quantum numbers ($\epsilon_1 \neq \epsilon_2$), the bi-particle ground state is a singlet state. The first excited state is a 4-fold degenerate state, with energy $E_1 = \epsilon_1 + \epsilon_2$ and four possible arrangements of the two spins. This degeneracy can be partially lifted in the presence of the small Zeeman field.

\begin{figure}[t]
\centering
\includegraphics[width=8.cm]{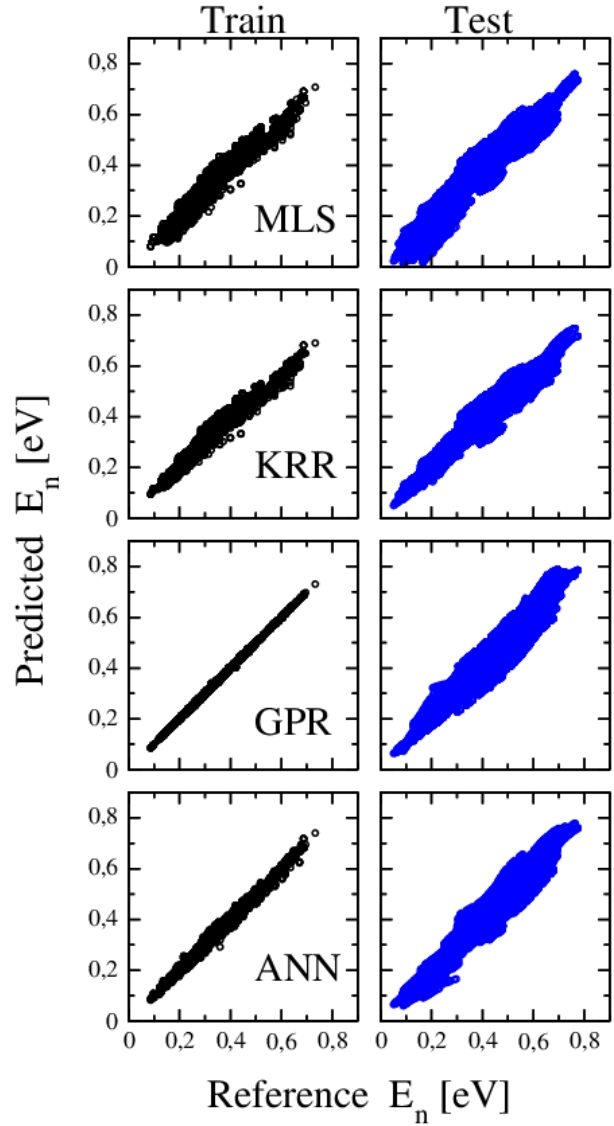} 
	\caption{Performance analysis of the ML methods (MLS, KRR, GPR, ANN), showing predicted vs. reference eigenvalues ($E_n$) for train (left column) and test (right column) sets. A number of $N_V \times N_{\rm int}$ non-equivalent systems are used for training, while for test, the equivalent systems are also added. The data sets contain $N_{\rm train} = 30 \times 3$ systems and $N_{\rm test} = 52 \times 16$ systems (plus all equivalent potentials). The prediction accuracies are determined by the $R^2$ coefficient of determination: MLS (0.85), KRR (0.91), GPR (0.92), ANN (0.96). For each system all 28 eigenvalues are plotted. The ANN mapping exhibits the best performance.}
\label{MLcompmethod}
\end{figure}

However, as the Coulomb interaction is included and increased, the energy levels are shifted towards higher values and the first excited state develops into a triplet state with total spin $S_z = \sum_i \bar{\sigma}_{z,i} = 1$. At the same time, the energy difference between the singlet and triplet states, $\Delta E = E_1 - E_0$, is getting smaller. While for ${\mathcal V_1}$ the energy difference $\Delta E$ is still visible for the largest interaction factor ($v_{\rm C}=1$), in the case of ${\mathcal V_2}$ a nearly 4-fold degenerate ground state appears.

This behavior is also captured in the evolution of the ground state charge density with $v_{\rm C}$ as depicted in Fig.\ \ref{chargedensity}. The main effect of the Coulomb interaction is the expansion of the electron charge in the local quantum well. Depending on the geometry of the confinement potential, the ground state can develop into a triplet state as it is the case for ${\mathcal V_2}$ potential, where two maxima become visible. In this case, taking into account the Zeeman field, the bi-particle state with total spin $S_z=1$ becomes the ground state and the first excited state has total spin $S_z=0$ (singlet state).

\begin{figure}[t]
\centering
\includegraphics[width=7.5cm]{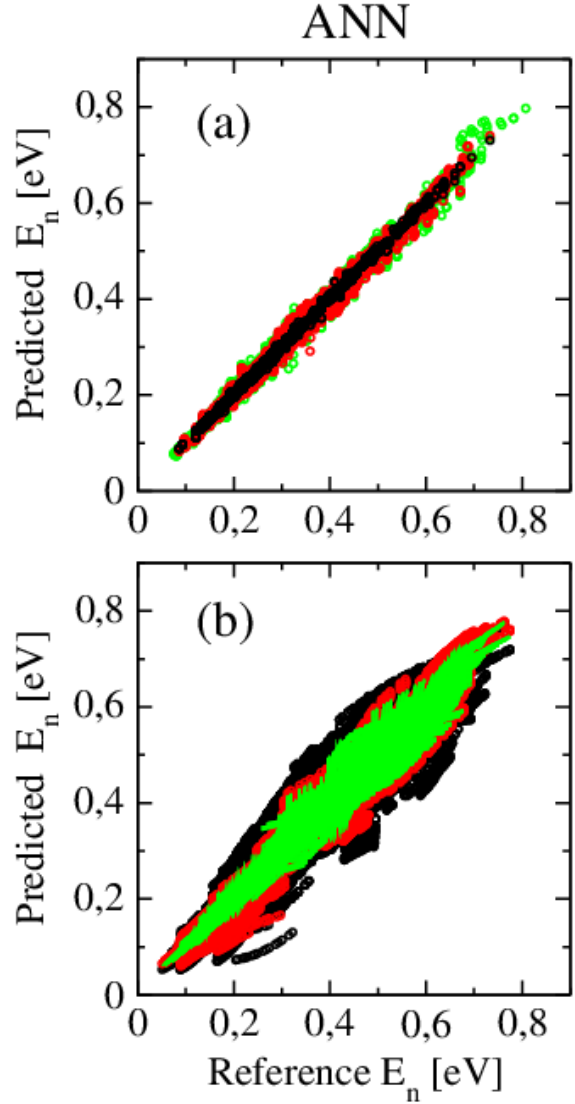} 
	\caption{The accuracies of ANN models for different sizes of the training sets: (a) train and (b) test sets. The $R^2$ values obtained for three different values $N_{\rm train} = N_V \times N_{\rm int}$, where the number of potentials $N_V=10$ ($R^2=0.81$) -- black colour, $N_V=30$ ($R^2=0.96$) -- red colour, $N_V=50$ ($R^2=0.97$) -- green colour and $N_{\rm int}=3$ values for the Coulomb interaction, $v_{\rm C}$. }  
\label{MLcomptrain}
\end{figure}

This transition can be followed from the $(E_n,v_{\rm C})$ maps shown in Fig.\ \ref{transition}(a,c). The blue arrow marks the singlet-triplet transition for ${\mathcal V_2}$, which is found for $v_{\rm C}\approx0.8$, as opposed to the case ${\mathcal V_1}$, where the first two eigenvalues do not cross. The expansion coefficients of the ground state in the 2-particle basis show a consistent behavior: a smooth dependence is obtained for ${\mathcal V_1}$, while a sharp discontinuities appear at the transition point for ${\mathcal V_2}$, as can be seen from Fig.\ \ref{transition}(b,d). The coefficient $C_{n0}=1$ for $v_{\rm C} = 0$, i.e. identical to the ground state of the non-interacting system, and for $v_{\rm C} > 0$ the ground state is a super-position of non-interacting 2-particle states. One should note that, in each case, the 2-particle basis elements are consistent with the total spin of the system. Denoting the 2-particle basis elements by 
$|\uparrow_0\downarrow_1\uparrow_2\downarrow_3\ldots\uparrow_6\downarrow_7\rangle$, 
one can easily check the total spin of the states entering the superposition indicated in Fig.\ \ref{transition}(d). One should also emphasize that these discontinuous contributions are smoothly matched by the coefficients of the 1st excited state, where the corresponding triplet-singlet transition can be observed.

\subsection{Prediction of the eigenvalues}

The full set of 512 potentials contains a number of equivalent systems, which can be obtained by rotations and mirroring, while the number of non-equivalent potentials is 102. In order to maximally exploit the training data, we included all the (computationally free) equivalent configurations. This should also lead the model to naturally incorporate the symmetry of the system. Similarly, the test set is assembled with all equivalent potentials sets, buts strictly distinct from the training set.

The train set contains $N_{\rm train} = N_V \times N_{\rm int}$ examples, where $N_V$ is the number of potentials and $N_{\rm int}$ is the number of selected Coulomb interactions for training. In order to test the accuracies of the different ML methods, we consider a reference calculation with $30 \times 3$ training examples and $52 \times 16$ test examples, plus all equivalent potentials added in the train/test sets. We used $v_{\rm C} = 0.1, 0.5, 0.9$ as training interaction strengths. The results are indicated in Fig.\ \ref{MLcompmethod}. The linear method (MLS) offers a reasonable accuracy ($R^2=0.85$), which is outperformed by the non-linear methods: KRR (0.91), GPR (0.92) and ANN (0.96). Another feature is that the non-linear methods offer a significantly better accuracy for lower ranked eigenvalues (e.g. ground state and first excited levels). Although the $R^2$ values are comparable, the ANN performs slightly better. On the other hand, in the current setup, GPR presents overfitting, which potentially limits the test accuracy. 

Another noticeable difference between the non-linear methods is that KRR and GPR learn the rotation and mirroring symmetries, while the ANN method just brings a very good approximation. This is due to the fact that KRR and GPR rely on exact optimization methods, as opposed to ANNs. While the ANN based method is most accurate for the current set-up, the training time is slightly larger than for KRR and GPR.  

Changing the number of selected interactions for training, $N_{\rm train} = 3, 4, 5$, hereby including $v_{\rm C} =0.3$ and $0.7$, produces little changes in the test accuracies, as the dependence of the eigenvalues on the interaction coordinate is almost linear. However, changing the number of potentials in the training set, $N_V = 10, 30, 50$ plus the respective equivalent potentials, produces visible effects, as shown in Fig.\ \ref{MLcomptrain}. As $N_V$ is getting larger, the train accuracy slightly diminishes, but the prediction for the test set improves: $R^2 = 0.81, 0.96, 0.97$.  

This shows that it is possible for this class of systems to predict the entire spectrum (28 eigenvalues) with quite high accuracy, having as input an $N_g+1$ feature vector, representing the binary representation of the device potential and the strength of the Coulomb interaction, $v_{\rm C}$. The non-linear methods, like KRR, GPR, ANN, bring $\sim$ 10\% improvement in the accuracy as measured by $R^2$, which may be important in the design of optoelectronic devices.   

\subsection{Prediction of the singlet-triplet transition}

Another important aspect concerns the existence of a transition point between the singlet and triplet states, as can be observed for the ground state in some systems when $v_{\rm C}$ is varied. Developing predictive methods that are able to identify the singlet-triplet transition has a two-fold implication: (i) from technological point of view, it supports system design where the total spin can be switched and (ii) from technical point of view, it can enlarge the interaction strength interval for predicting the eigenvectors. 

After testing several classification algorithms for identifying the discontinuity in the first eigenvector, it was found that none outperformed simple linear classification, being in the same range of accuracies. The deciding feature of the linear algorithm was established to be the presence or absence of the central potential block. This simple rule, inferred by the classification method, is further used to detect the presence of singlet-triplet transitions. 

As such, 83.59\% of the systems are correctly classified. More precisely, following this simple rule one can perform a prediction with: 2.15\% false negative (11 cases), 14.26\% false positive (73 cases), 47.85\% true positive (245 cases) and 35.74\% true negative (183 cases). If the symmetry of the 9 cell system is taken into account, one would have a total of 102 unique potentials, out of which 79.41\% are correctly classified (81 cases) and 20.58\% are misclassified (21 cases).

\section{Conclusions}

We defined a large class of many-electron problems based on a quantum dot system, controlled by a top-gate array. The binary-valued voltages applied on the gates, as an external input, and the strength of the Coulomb interaction, as a material related property, determine the energy spectra and singlet/triplet nature of the ground state. Here, we considered bi-particle problems, which we have solved by high throughput calculations using the exact diagonalization technique. From technological point of view, it is important to have accurate overview concerning the optoelectronic switching properties of the quantum system, which are mainly reflected by the lower part of the energy spectrum and total spin of the respective quantum states. As the number of candidate systems grows exponentially with the number of gates, we investigate in how far high throughput exact diagonalization calculations combined with ML techniques are able to accurately reproduce the energy spectra. We employ multi-target regression methods, which rely on readily available information as feature vectors: the binary representation of the potential configurations and the strength of the Coulomb interaction. The non-linear methods like KRR, GPR and ANN reproduce quite well the reference values, with $R^2$ coefficients larger than $0.9$. These methods outperform the linear regression models, based on MLS, particularly concerning the low lying states. Furthermore, the existence of a singlet-triplet transition for the ground state can be determined with a reasonable accuracy ($R^2=84\%$) by identifying a simple rule, obtained from a linear classifier. Our results show that the ML techniques can significantly reduce the computation burden of exact calculations, being able to reasonably predict the reference values. This approach can aid the design process of novel quantum devices, while the methods can be extended to systems where an exhaustive computation of the many-body states becomes unfeasible.\\

{\bf Acknowledgments} \\

This work was supported by a grant of the Romanian Ministry of Research, Innovation and Digitalization, CNCS - UEFISCDI, project number PN-III-P4-ID-PCE-2020-1142, within PNCDI III. We acknowledge useful discussions with V.V. Baran and D.V. Anghel.\\

{\bf References}\\

\bibliography{manuscript_R1.bbl} 
\bibliographystyle{iopart-num}

\end{document}